\documentclass[a4paper,11pt]{article}
\usepackage{graphicx}
\usepackage{dcolumn}
\usepackage{bm}
\usepackage{epsfig,amsmath}
\usepackage{amssymb}
\usepackage{subcaption}
\usepackage[normalem]{ulem}
\usepackage[usenames,dvipsnames]{color}
\usepackage[pagebackref=false, colorlinks=true]{hyperref}
\definecolor{redish}{rgb}{0.7,0.2,0.0}  
\definecolor{bluish}{rgb}{0.2,0.5,0.8}

\hypersetup{linkcolor=redish,          
                  citecolor=blue,        
                  filecolor=magenta,      
                  urlcolor=bluish}          

\DeclareFontFamily{U}{rsfs}{}         
\DeclareFontShape{U}{rsfs}{m}{n}{<5> rsfs5 <6><7> rsfs7          %
  <8><9><10><10.95><12><14.4><17.28><20.74><24.88> rsfs10}{}     %
\DeclareMathAlphabet{\mathfs}{U}{rsfs}{m}{n}

\title{Probing (sub-)solar-mass black holes and superspinars with current and next-generation gravitational wave observatories}

\author{
K. S. Sruthy$^1$\footnote{sruthy.mcnsmpl2023@learner.manipal.edu},
N. V. Krishnendu$^2$\footnote{k.naderivarium@bham.ac.uk},
Chandrachur Chakraborty$^1$\footnote{chandrachur.c@manipal.edu},
~and~ Nami Uchikata$^3$\footnote{uchikata@icrr.u-tokyo.ac.jp}
\\
{$^1$\it Manipal Centre for Natural Sciences, Manipal Academy of Higher Education,}\\
{\it Manipal 576104, India}
\\
{$^2$\it School of Physics and Astronomy, University of Birmingham,}\\
{\it Edgbaston, Birmingham, B15 2TT, UK}
\\
{$^3$\it Institute for Cosmic Ray Research, University of Tokyo,}\\
{\it Kashiwa City, Chiba 277-8582, Japan}
}

\date{}

\begin{document}

\maketitle

\begin{abstract}
We investigate the detectability and parameter estimation prospects of (sub-)solar-mass collapsed binaries with current and next-generation gravitational wave detectors, allowing the dimensionless component spins to extend phenomenologically beyond the Kerr bound, \(|\chi|>1\) (where $\chi$ is the spin parameter). Using the inspiral-only TaylorF2 waveform and a Fisher-matrix analysis, we study binaries with component masses in the range \(0.1\!-\!2\,M_\odot\) and compare the performance of Advanced LIGO with that of a third-generation network consisting of Cosmic Explorer and the Einstein Telescope. We find that third-generation detectors substantially improve the signal-to-noise ratio and the measurement of intrinsic binary parameters, particularly the component spins, while also providing significantly improved sky localization. For favorable configurations, the primary spin can be measured with precision \(\Delta\chi_{1z}\sim10^{-3}-10^{-2}\), demonstrating enhanced sensitivity to potential Kerr-bound violations. However, strong anti-correlations between the component spins limit the interpretation of small marginalized spin uncertainties as definitive evidence for superspinars. We also examine formation scenarios for (sub-)solar-mass collapsed binaries and show that dense dark-matter environments, particularly the Galactic Centre, can significantly enhance their formation and merger rates. Our results demonstrate the potential of future gravitational wave observations to probe low-mass collapsed objects and possible departures from the Kerr bound, while highlighting the need for physically consistent superspinar waveform models and Bayesian model comparison for a definitive interpretation.
\end{abstract}

\section{\label{sec:Intro} Introduction}
The direct detection of gravitational waves (GWs) from compact binary mergers has inaugurated a new era in observational gravitational physics, providing an unprecedented opportunity to probe the nature of compact objects and test general relativity in the strong-field regime. Since the first observation by the LIGO and Virgo collaborations in 2015 \cite{Abbott_2016}, the rapidly growing catalog of GW events has revealed a diverse population of black holes and neutron stars spanning a wide range of masses and spins. The latest GW transient catalog, GWTC-5.0 \cite{ligo_2026,ligo_2026a,ligo_2026b}, contains a large number of compact-binary mergers detected by the global detector network, substantially expanding the observed population of compact objects and enabling increasingly precise tests of gravity in the strong-field regime \cite{LIGOScientific:2026qni,LIGOScientific:2026fcf,LIGOScientific:2026wpt,LIGOScientific:2025rid,ligo_2026_testofgr}. These observations have significantly advanced our understanding of compact-object astrophysics and the behavior of gravity under extreme conditions.

As the GW catalog has expanded, the explored mass range of compact objects has extended toward both the lower and upper mass gaps. Standard stellar evolution predicts that black holes cannot form with masses below approximately $3M_{\odot}$. However, several candidate events have been reported with component masses potentially in the sub-solar regime. For example, the candidate event SSM200308 has been reported with component masses of approximately $0.62^{+0.46}_{-0.20}\,M_\odot$ and $0.27^{+0.12}_{-0.10}\,M_\odot$ \cite{Prunier_2024}, while one of the binary candidates of the event SSM170401 is reported as $0.76^{+0.50}_{-0.14}\,M_\odot$ \cite{Morr_s_2023}. Another recent event S251112cm has a chirp mass in the range $0.1$--$0.87\,M_\odot$ \cite{Magaraggia_2026}. In addition, the events GW190425 \cite{Abbott_2020_a} and GW190814 \cite{Abbott_2020} could involve near-solar-mass objects, whose nature---either a heavy neutron star or a light black hole---remains uncertain \cite{Dasgupta_2021}. The search for binary mergers with atleast one sub-solar-mass components in the first part of fourth run of LVK network, and its parameter estimations studies has done by \cite{abac_2026} and \cite{christensen_2026} respectively. A confirmed detection of a (sub-)solar-mass black hole would therefore point toward non-standard formation channels and would have important implications for compact-object astrophysics.

At present, no widely accepted astrophysical mechanism is known to produce
(sub-)solar-mass black holes through conventional stellar evolution. A leading
alternative is the primordial formation of such objects through the collapse of
overdensities in the early Universe. Proposed mechanisms include enhanced
inflationary density perturbations \cite{Bellido_1996}, the collapse of cosmic-string
loops \cite{Hawking_1989}, and early-Universe phase transitions
\cite{Hawking_1982}. More recently, it has been suggested that the accumulation
of non-annihilating dark matter inside compact stars could trigger their
transmutation into low-mass black holes \cite{Dasgupta_2021}. Related mechanisms may also lead to over-spinning collapsed objects \cite{Chakraborty_2024,Chakraborty_2024b}, commonly referred to as Kerr superspinars. Although
these formation scenarios remain speculative, they motivate the study of the
GW signatures of low-mass collapsed objects across both the black hole ($0<|\chi|\leq1$, where $\chi$ is the dimensionless spin parameter) and superspinar ($|\chi| > 1$) spin regimes. In particular, it is important to
determine  whether such objects can form through astrophysically permitted channels, whether mergers involving such objects would be detectable and
whether their signals could be distinguished from those of conventional Kerr
black holes. GW detections provide a direct and independent avenue for testing possible violations of the Kerr bound and, more broadly, the cosmic censorship conjecture \cite{Wade_2013}.

Despite the theoretical possibilities, the observational implications of (sub-)solar-mass black holes and superspinars remain largely unexplored. In particular, their GW signatures have also not been systematically studied in the context of potential violations of the Kerr bound (i.e., $|\chi| > 1$). Most waveform models currently employed in GW data analysis assume the validity of the Kerr bound and are calibrated within the regime $0< |\chi| < 1$. Moreover, the uncertainties in spin measurements with current second-generation detectors are often sufficiently large that distinguishing between an extremely rapidly rotating black hole and a superspinar is challenging. Future third-generation detectors, such as the Cosmic Explorer (CE) and the Einstein Telescope (ET), are expected to achieve significantly improved strain sensitivity and substantially enhanced parameter estimation precision across a broad redshift range \cite{Punturo_2010, Hild_2011, Reitze_2019}.

Recent studies have explored the detectability and characterization of low-mass compact binaries with current and future GW observatories. These investigations have examined the detectability \cite{giri_2025} and the measurability of masses, spins, and sky locations for sub-solar-mass binaries \cite{Wolfe_2023}, parameter estimation prospects with third-generation detectors \cite{Begnoni_2026}, and the classification of (sub-)solar collapsed objects through their tidal and GW signatures \cite{Golomb_2024}. Related works have also demonstrated the potential of future observatories such as NEMO (Neutron Star Extreme Matter Observatory), CE, and ET to distinguish low-mass black holes from neutron stars, probe dark-matter-induced transmutation scenarios and constrain dark matter through the non-detection of transmuted black hole mergers \cite{bhattacharya_2025distinguishing,bhattacharya_2023can,Crescimbeni:2024cwh,crescimbeni2025cosmology}. Collectively, these studies have established (sub-)solar-mass collapsed binaries as a promising target for current and future GW observatories.

However, most previous studies have assumed the validity of the Kerr bound and restricted the component spins to the black hole regime, $|\chi| < 1$. In this work, we investigate the extent to which current and next-generation GW observatories can probe phenomenological extensions of the spin parameter beyond the Kerr bound in the (sub-)solar-mass regime. Rather than constructing a complete waveform model for superspinars, we assess the sensitivity of inspiral-phase GW observations to apparent Kerr-bound violations and quantify the corresponding parameter estimation capabilities of second- and third-generation detector networks. To this end, we construct astrophysically motivated representative populations of (sub-)solar-mass collapsed binaries, model their inspiral signals using the TaylorF2 waveform approximant following \cite{Uchikata_2021}, and employ Fisher-matrix techniques to estimate the precision with which key source parameters, particularly the component masses and spins, can be measured. We further examine the regions of parameter space in which future detectors may be capable of placing meaningful constraints on apparent violations of the Kerr bound. In addition, we investigate several representative formation scenarios for (sub-)solar-mass collapsed binaries and assess their implications for the expected merger rates and observational prospects of such systems.

Possible observational differences between binary black hole and superspinar mergers may also arise during the post-merger ringdown phase, especially if the remnant violates the Kerr bound. However, the ringdown spectrum of superspinars depends sensitively on the assumed boundary conditions at the surface or interior of the object, which remain uncertain, and no complete inspiral--merger--ringdown waveform models are currently available. We therefore focus on the inspiral regime, where the GW signal can be studied in a comparatively robust and model-independent manner. We construct representative populations of (sub-)solar-mass binaries and model their GW signals during the inspiral phase. Using Fisher-matrix techniques, we estimate the precision with which key source parameters, particularly the masses and spin parameters, can be determined with second- and third-generation detectors. We additionally examine several proposed formation scenarios and estimate the corresponding merger rates under specific assumptions, providing an astrophysical context for the detectability study. In addition to this, we examine the sky-localization prospects relevant for subsequent multimessenger follow-up.

The paper is organized as follows. We briefly discuss the Kerr black holes and superspinars in section \ref{sec:kbs}, and the possible formation channels of (sub-)solar-mass black holes and superspinars in section~\ref{sec:SSMBH_NS}. The methodology employed in this work is briefly outlined in section~\ref{sec:methodology}.  The details of the binary population studied in this work are described in section~\ref{binarypopulation}.
In section~\ref{sec:Network}, we describe the GW detector network configurations considered in this work. The parameter estimation framework, including the simulated binary population, signal-to-noise ratio calculation, Fisher matrix formalism, waveform model, and parameter space, is presented in section~\ref{sec:para_est}. The results of our analysis and their implications for the detectability of low-mass binaries as well as Kerr-bound violations with current and future detector networks are presented in section~\ref{sec:Results}. In section~\ref{sec:formation-scenario}, we examine astrophysical binary formation scenarios and assess the expected merger rates and detectability of such systems in different environments. Finally, in section~\ref{sec:Summary} we summarize our findings and discuss their implications for tests of strong-field gravity and the possible existence of (sub-)solar-mass binaries.

\section{Kerr black holes and Kerr superspinars\label{sec:kbs}}

Within general relativity, the spacetime of an isolated, stationary, axisymmetric collapsed object is uniquely described by the Kerr solution \cite{Kerr_R_P_1963}. In this framework the geometry is completely determined by two parameters: the mass $M$ and the angular momentum $J$. The dimensionless spin parameter is defined as $\chi = cJ/GM^2$ where $c$ is the speed of light in vacuum and $G$ is the gravitational constant. The radius of the event horizon in Kerr spacetime is given by $R_{\rm h}=GM/c^2 \left(1+\sqrt{1-\chi^2}\right)$. For $|\chi|\leq1$, the horizon radius is real, and the Kerr solution represents a black hole whose central singularity is hidden behind the event horizon. By contrast, when $|\chi|>1$, the horizon radius is no longer real, and the Kerr geometry becomes super-extremal and horizonless. Such configurations are commonly referred to as Kerr naked singularities \cite{Chakraborty_2017a,Chakraborty_2017b} or, Kerr superspinars \cite{Gimon_2009, cpa, Stuchlik_2010}.

The Kerr bound is closely related to the cosmic censorship conjecture, which
posits that singularities formed in gravitational collapse are generically hidden
from distant observers. Testing the validity of this conjecture remains one of
the fundamental problems in gravitational physics. Superspinars have been
explored in various theoretical contexts, including string-inspired extensions of
gravity \cite{Gimon_2009}, and have been proposed as potential sources of high-
energy astrophysical phenomena due to their enhanced efficiency for energy
extraction \cite{Patil_2016, Chakraborty_2024}. It has also been argued that, under certain assumptions, horizonless superspinning objects may remain consistent with existing electromagnetic observations of sources such as M87* \cite{Bambi_2019,GhasemiNodehi2021M87} and GRO J1655--40 \cite{CCGMmonople}. Nevertheless, all  GW observations to date remain consistent with the Kerr bound,
although the spin parameters of many events are still relatively weakly
constrained.

\subsection{\label{sec:SSMBH_NS}Formation channels of (sub-)solar-mass Kerr black holes and superspinars}

Several candidate collapsed objects (see section \ref{sec:Intro}) identified in recent GW observations appear to have masses comparable to those of neutron stars and white dwarfs. If confirmed, such objects would challenge the conventional picture of collapsed-object formation and motivate the exploration of alternative astrophysical and cosmological channels. In particular, the formation of spinning collapsed objects in the (sub-)solar-mass regime is difficult to accommodate within standard stellar evolution, which generally does not produce collapsed objects substantially below a few solar masses. Here, we briefly discuss possible non-standard formation channels for such low-mass collapsed objects.

\subsubsection{Dark matter induced collapse of compact stars \label{ss1}}
 An alternative formation pathway involves the interaction between compact stars and dark matter. Compact objects such as white dwarfs or neutron stars can gravitationally capture dark-matter particles, which accumulate in their interiors through repeated scattering interactions. If the captured dark matter is asymmetric or non-self-annihilating, it can form a dense self-gravitating core that eventually undergoes gravitational collapse, producing a tiny black hole at the center of the host star (see section 2 of \cite{Chakraborty_2024b} for detailed formation criteria). The subsequent accretion of the surrounding stellar material can convert the original compact star into a low-mass black hole, a process often referred to as `transmutation' \cite{mcd2012}.

Recent studies have investigated the collapse of dark-matter cores inside spinning white dwarfs and shown that this process can produce (sub-)solar-mass collapsed objects, including black holes and superspinars, under certain conditions \cite{Chakraborty_2024, Chakraborty_2024b}. The outcome depends on the properties of the host star—particularly its rotation \cite{acb1}, viscosity \cite{acb1}, and magnetic fields \cite{acb2}—as well as on the microphysical properties \cite{acb3} of the dark matter.
The fate of the system is largely determined by the spin of the white dwarf. For slow rotation, the tiny black hole formed at the center grows efficiently through accretion and eventually consumes the star, producing a (sub-)solar-mass black hole depending on the progenitor's mass. For intermediate spins, the collapse may occur without the formation of an event horizon, potentially resulting in a naked singularity \cite{Chakraborty_2017b, Chakraborty_2024} or superspinar \cite{Gimon_2009}. In contrast, rapid rotation can suppress accretion through centrifugal support and viscous effects, leading to a stalled configuration in which the collapsed remnant remains embedded within the star as an `endoparasitic' black hole \cite{acb1, acb2}.
Because these processes occur in rotating stellar environments, the resulting remnants may inherit substantial angular momentum. In sufficiently rotating systems, this may lead to ultra-spinning collapsed objects that exceed the Kerr bound, providing a possible astrophysical pathway for the formation of superspinars \cite{Chakraborty_2024, acb1}.
The situation is different for neutron stars. Although dark-matter-induced collapse can also produce low-mass black holes in neutron-star systems \cite{Chakraborty_2024, acb1, East_2019}, their available angular momentum is generally insufficient to produce ultra-spinning remnants. Consequently, the formation of superspinars through this channel is expected to be significantly less likely  \cite{Chakraborty_2024} than in white-dwarf progenitors.

\subsubsection{Primordial black holes \label{ss2}}
One of the most widely discussed possibilities is the formation of primordial black holes (PBHs) in the early Universe. In this scenario, black holes form from the collapse of large primordial density fluctuations shortly after the Big Bang. The mass of a PBH is determined by the cosmological horizon mass at the time of formation, allowing PBHs to span an extremely broad mass spectrum ranging from very small masses ($\sim 10^{11}\,\mathrm{kg}$ \cite{page_1976} or even lower \cite{chakraborty_2022}) to stellar and supermassive scales. Several mechanisms have been proposed to generate the required overdensities, including enhanced inflationary fluctuations, collapse of cosmic-string loops, and first-order phase transitions in the early Universe.
Although PBHs can naturally populate the (sub-)solar-mass regime, a few studies \cite{mirbabayi_2020, luca_2020} indicate that directly formed PBHs could possess small dimensionless spins (but see section \ref{ss4}). However, this could be modified by subsequent astrophysical processes.

\subsubsection{Primordial black holes captured by compact stars\label{ss3}}
A recently proposed indirect-PBH scenario provides a possible connection between primordial seeds and astrophysically transmuted low-mass remnants
\cite{BS26}. In this scenario, a  with mass
$M_{\rm PBH}\ll M_{*}$ is captured by a compact star (e.g., white dwarfs \cite{capwd, Carr2020} or neutron stars \cite{capns, Carr2020}) of mass $M_{*}$ and subsequently settles toward its centre. Immediately after that, the captured PBH can act as a seed black hole and grow through accretion of the surrounding stellar material \cite{Carr2020}. The subsequent evolution is qualitatively similar to the dark-matter induced transmutation scenarios discussed above (see section \ref{ss1}), with the properties of the final remnant determined primarily by the mass and rotational state of the host star. Consequently, this channel may also produce (sub-)solar-mass spinning black holes and, under suitable conditions, superspinars \cite{Chakraborty_2024, Chakraborty_2024b, acb1} (and references therein), depending on the mass and spin of the progenitor star. Unlike the direct-PBH scenario, the mass and spin of the final remnant is therefore set primarily by the host star rather than by the primordial seed. Estimates indicate that this mechanism may produce appreciable
populations of (sub-)solar-mass collapsed objects \cite{BS26}.

\subsubsection{Spin evolution of primordial black holes and primordial origin of superspinars \label{ss4}}

A more speculative possibility is that ultra-spinning collapsed objects formed directly in the early Universe. PBHs produced from nearly isotropic density perturbations during radiation domination are generally expected to have small spins \cite{mirbabayi_2020,luca_2020}, whereas formation during an early matter-dominated epoch can yield rapidly rotating PBHs with dimensionless spin approaching the Kerr bound \cite{Harada_2017,deJong_2023}.

Note that the direct formation of primordial superspinars requires physics beyond the conventional PBH-collapse scenario. Such objects may have originated as remnants of the high-energy early Universe, where string-scale or quantum-gravity corrections could invalidate the classical Kerr bound and permit collapsed objects with $\chi>1$. Violations of the Kerr bound have therefore been proposed as possible signatures of string theory \cite{Gimon_2009}. Regular or quantum-corrected collapsed-object models may further replace the Kerr singularity with a nonsingular core, allowing an over-spinning configuration without an event horizon or an exposed curvature singularity \cite{Li_Bambi_2013,Yang_2022}. The astrophysical appearance and accretion properties of the primordial superspinars were investigated in \cite{Stuchlik_2012}. Thus, primordial superspinars could, in principle, populate the (sub-)solar-mass regime, although their formation and survival depend essentially on non-standard gravitational or quantum-gravity physics.

The different formation channels discussed above
can lead to distinct distributions of masses, spins,
and merger environments for collapsed binaries. Primordial formation mechanisms typically operate in
the early Universe and are expected to produce collapsed objects with a broad mass spectrum and spin \cite{Stuchlik_2012, Harada_2017, Li_Bambi_2013,Yang_2022, deJong_2023, mirbabayi_2020,luca_2020}, whereas
compact-star transmutation (induced by dark matter \cite{mcd2012, Chakraborty_2024, acb1} or PBH \cite{capns, capwd, Carr2020}) scenarios can yield remnants with masses comparable to those of neutron
stars or white dwarfs and potentially substantial angular momentum. GW observations of (sub-)solar-mass collapsed binaries therefore provide a promising
avenue for probing these formation pathways and testing the possible existence of rapidly spinning collapsed objects such as superspinars.

While several theoretical scenarios permit the formation of superspinars discussed above, their abundance and merger rates remain uncertain. Consequently, in this work we do not attempt to predict a superspinar population, but instead investigate the observational question of how well current and future GW detectors could constrain such objects if they exist in collapsed binaries.

\section{\label{sec:methodology} Methodology}

\subsection{Details of the binary population}\label{binarypopulation}
We simulate a population of $\sim 1000$ (sub-)solar-mass collapsed binaries to investigate parameter estimation uncertainties with current and next-generation detector networks. The population includes both Kerr black holes ($\chi \leq 1$) and Kerr superspinars ($\chi >1$), by allowing the dimensionless spin parameter $\chi$ to extend beyond the Kerr bound. Each binary is characterized by the following parameters, drawn from the distributions described below:

\begin{itemize}

\item \textit{Component masses:}
The primary and secondary masses $m_1$ and $m_2$ are drawn independently from a uniform distribution in the range $$0.1~M_\odot \leq m_{1,2} \leq 2~M_\odot,$$
such that $m_1>m_2$.

\item \textit{Component spins:}
The dimensionless spin projections along the orbital angular momentum axis, $\chi_{1z}$ and $\chi_{2z}$, are drawn uniformly from the range $$-1.5 \leq \chi_{1z,2z} \leq 1.5.$$
This choice is motivated by the following considerations.

First, the primary objective of this work is to assess the detectability and parameter estimation prospects of (sub-)solar-mass collapsed binaries, including the possibility of detectability of potential violations of the Kerr bound. We therefore extend the allowed spin range beyond the classical limit $\chi=1$ in order to include superspinar configurations within a phenomenological framework.

Second, we do the analysis to aligned and anti-aligned spin systems.

Third, the inspiral waveform model employed in this work (TaylorF2) incorporates spin effects through post-Newtonian phase corrections and does not explicitly enforce the Kerr bound. For moderate extensions beyond unity, the waveform remains mathematically well-defined. We therefore adopt an upper limit of $|\chi|=1.5$ \cite{Uchikata_2021}, which allows us to probe Kerr-bound violations without entering extreme spin regimes where post-Newtonian expansions may become unreliable.

Finally, we emphasize that the adopted spin distribution is not intended to represent a realistic astrophysical prior. Rather, it is chosen as a phenomenological extension of the Kerr parameter space to quantify the sensitivity of current and future GW detector networks to potential Kerr-bound violations.

Additionally, we retain only systems with non-negative effective spin, $\chi_{\rm eff} = (m_1\chi_{1z} + m_2\chi_{2z})/(m_1+m_2) \ge 0$.

\item \textit{Luminosity distance:}
The sources are distributed uniformly in comoving volume within the luminosity-distance range \(1\,{\rm Mpc}\le D_L\le300\,{\rm Mpc}\). The lower bound is a practical boundary of the simulated population rather than a physical exclusion of closer sources, while the upper bound is chosen to explore the detection and parameter estimation prospects of these low-mass binaries over a local volume accessible to the detector networks considered. For each source, the corresponding redshift is calculated using the Planck 2018 cosmological parameters \cite{Planck2020Cosmological} as implemented in Astropy\cite{astropy:2013, astropy:2018, astropy:2022}. This distance range allows potential (sub-)solar-mass binaries to be detectable by both current and next-generation detectors. Because of the low masses considered here, the detector reach for these systems is limited to a few hundred Mpc even for third-generation networks.

\item \textit{Binary orientation:}
The inclination angle $\iota$ is sampled such that $\rm cos~\iota$ is uniformly distributed in [-1,1], ensuring isotropic orientation.

The right ascension $\alpha$ is drawn uniformly in [0, 2$\rm \pi$].
The declination $\delta$ is sampled isotropically by drawing $\rm sin~\delta$ uniformly in [-1,1].

The polarization angle $\psi$ is drawn uniformly in [0, $\rm \pi$].

The coalescence time $t_c$ and phase $\phi_c$ are fixed to zero without loss of generality, as they do not affect intrinsic parameter uncertainties in the Fisher-matrix framework.

\end{itemize}

These constitute the injection parameters used for our population-based parameter estimation study. The detectability of spin values exceeding the Kerr bound (i. e., $|\chi| > 1$) has been investigated in \cite{Uchikata_2021}, where it was shown that GW observations can constrain superspinars through precise measurements of the inspiral phasing. They further demonstrated that if a superspinar signal is analyzed using standard black hole waveform models that impose the Kerr bound ($|\chi| \leq 1$), the recovered source parameters become systematically biased, leading to incorrect estimates of both the masses and spins of the binary components.

\subsection{\label{sec:Network} Gravitational Wave Detector Network Configurations}

In this section, we describe the GW detector networks considered in our analysis. We examine two classes of networks: (i) a second-generation network consisting of the currently operating Advanced LIGO detectors, and (ii) a third-generation network composed of next-generation observatories such as CE and ET. The network configurations and naming conventions adopted in this work follow those introduced in \cite{Gupta_2024}.

\subsubsection{Second-Generation Detector Network}

We consider a two-detector network consisting of the Advanced LIGO observatories at Hanford (LHO) and Livingston (LLO), hereafter referred to as the HL network. The detectors are modeled using their design sensitivity power spectral densities. Within the GWBENCH framework \cite{Borhanian_2021}, these are denoted as aLIGO\_H and aLIGO\_L.

\subsubsection{Third-Generation Detector Network}

For the third-generation configuration, we consider a three-detector network composed of two CE interferometers and one ET.
The CE detectors are assumed to be L-shaped interferometers with 40 km arm length (CE-40) \cite{Reitze_2019}. Since their final locations are not yet fixed, we place them at the current LIGO Hanford and Livingston sites with the same orientations. These are denoted as CE-40\_H and CE-40\_L in GWBENCH \cite{Borhanian_2021}.
The ET is modeled as a triangular interferometer with three 10 km arms in a $60^\circ$ configuration \cite{Punturo_2010}. For concreteness, we place it at the Virgo site and denote it as ET\_V.
The networks analyzed in this work are summarized in Table~\ref{table1}.

\begin{table}
\centering
\begin{tabular}{l c c}
\hline
Network & Number of & Detectors  \\
Name &  XG Observatories & in Network\\
\hline
HL & 0 (2G only) & aLIGO\_H, aLIGO\_L \\
4040ET & 3 (XG) & CE-40\_H, CE-40\_L, ET\_V \\
\hline
\end{tabular}
\caption{Detector networks considered in this work. XG denotes next generation detector \cite{Gupta_2024}.}
\label{table1}
\end{table}

\subsection{\label{sec:para_est}Parameter Estimation Framework}

\subsubsection{\label{subsec:para_est}Signal-to-Noise Ratio and Detection Criteria}

Having described the detector network configurations and the simulated binary population, we now outline the parameter estimation framework employed in this work. The detectability of a GW signal and the precision with which its parameters can be measured depend primarily on the signal-to-noise ratio (SNR), $\rho$. For a single detector $A$, the optimal SNR is defined as \cite{Gupta_2024, Cutler_1994}:

\begin{equation}
\rho_A^2 = 4 \int_{f_{\text{low}}}^{f_{\text{upper}}}
\frac{|\tilde{h}_A(f)|^2}{S_n^A(f)} \, df,
\end{equation}
where $\tilde{h}_A$ is the frequency-domain waveform at detector $A$, $S_n^A(f)$ is the one-sided noise power spectral density (PSD) of that detector, and $f_{\rm low}$ and $f_{\rm upper}$ denote the lower and upper cutoff frequencies, respectively. For a network of detectors, the total SNR is obtained by summing in quadrature:

\begin{equation}
\rho_{\rm net} = \left( \sum_A \rho_A^2 \right)^{1/2}.
\end{equation}
An event is considered detectable if the network SNR satisfies $\rho \geq 8$. This threshold is commonly adopted in GW detectability studies to approximate the sensitivity of matched-filter searches \cite{Cutler_1994}. However, the Fisher-matrix approximation is reliable only in the high SNR regime. Therefore, in the parameter estimation analysis presented in this work, we restrict our study to systems with $\rho \geq 10$.

\subsubsection{\label{subsec:fisher}Fisher Matrix Formalism}

To estimate parameter uncertainties, we employ the Fisher information matrix formalism. For a given detector, the Fisher information matrix $\Gamma_{ab}$ which is related to the derivatives of the waveform with respect to the set of source parameters $\lambda$ is defined as:
\begin{equation}
    \Gamma_{ab} = 2\int_{f_{\rm low}}^{f_{upper}}\frac{\tilde{h}_{A,a}\tilde{h}_{A,b}^*+\tilde{h}_{A,a}^*\tilde{h}_{A,b}}{S_n^A}df,
\end{equation}
where $*$ corresponds to complex conjugation and comma denotes the differentiation with respect to various elements of the parameter space $\lambda$. The inverse of $\Gamma_{ab}$ is the covariance matrix $\Sigma_{ab}$, and the square root of diagonal elements of the covariance matrix gives the $1\sigma$ uncertainty range of parameter measurements for a given detector:
\begin{equation}
    \sigma_a = \sqrt{\Sigma_{aa}}.
\end{equation}
Fisher-matrix techniques remain widely used for forecasting the parameter estimation capabilities of future GW detector networks, particularly in studies involving large source populations and third-generation observatories \cite{Cutler_1994,Iacovelli_2022,Gupta_2024,Arun_2005a}. Fisher-matrix estimates are known to become optimistic in highly correlated parameter spaces and do not account for waveform systematics \cite{Vallisneri_2008}. The reported uncertainties should therefore be interpreted as approximate high-SNR forecasts. We also note that the exceptionally long inspiral duration of (sub-)solar-mass binaries makes comprehensive Bayesian analyses computationally expensive across the large parameter space explored in this work. Thus, the Fisher information matrix remains a standard tool for preliminary forecasts \cite{Rodriguez_2013} with next-generation detectors. Previous studies have nevertheless found that, in the high-SNR regime, Fisher-matrix predictions generally recover the same order-of-magnitude parameter uncertainties as full Bayesian analyses, making them a useful tool for assessing the parameter estimation prospects of future detectors \cite{Vallisneri_2011, Rodriguez_2013, Cutler_1994, Vallisneri_2008}.

All Fisher-matrix calculations are performed using the publicly available GWBENCH package \cite{Borhanian_2021}, which interfaces with waveform models implemented in the LIGO Algorithms Library (LAL) \cite{lalsuite}. The parameter space $\lambda$ is defined as:
\begin{equation}
    \lambda = \{\mathcal{M}_c,\eta,\chi_{1z},\chi_{2z},D_L,\iota,\alpha,\delta,\psi,t_c,\phi_c\},
\end{equation}
where $\mathcal{M}_c = \frac{(m_1 m_2)^\frac{3}{5}} {(m_1 + m_2)^\frac{1}{5}}$ is the chirp mass, $\eta =\frac{m_1 m_2} {(m_1 + m_2)^2}$ the symmetric mass ratio, $\chi_{1z}$ and $\chi_{2z}$ the spin projections along the orbital angular momentum axis, $D_L$ the luminosity distance, $\iota$ the inclination angle, ($\alpha~,~\delta$) the sky location, $\psi$ the polarization angle, and $t_c$ and $\phi_c$ the coalescence time and phase (see section~\ref{sec:SSMBH_NS} for more details.).

The lower cutoff frequency of the waveform is taken to be $f_{\rm low} = 20$~Hz for HL network, consistent with the conventional low-frequency limit used for Advanced LIGO design-sensitivity studies, and $f_{\rm low} = 5$~Hz for 4040ET network to represent the extended low-frequency sensitivity assumed for the third-generation configuration.

The lower cutoff frequency $f_{\rm low}$ of the waveform is set by the low-frequency sensitivity of the detectors, not by astrophysical considerations. Below $f_{\rm low}$ the detector noise rises steeply and the signal contributes negligibly to the SNR. Third-generation detectors are designed to have a much lower seismic and Newtonian noise limited cutoff. The improved low-frequency sensitivity of these detectors means that a much longer part of the inspiral is observable, which increases the frequency interval over which the signal contributes to the SNR and to the Fisher information. This is particularly important for (sub-)solar-mass binaries, whose inspiral signals are long-lived and spend a large number of cycles at low frequencies.

Since the waveform model used here describes only the inspiral phase, we truncate the signal at the GW frequency corresponding to the innermost stable circular orbit (ISCO), $f_{\rm ISCO}$, in order to avoid contamination from the post-inspiral regime \cite{Cutler_1994}.

Since the present analysis includes superspinning configurations with $|\chi| > 1$, the standard Kerr ISCO expressions \cite{Favata} require careful interpretation and need to be revisited beyond the extremal spin limit. We therefore use the Schwarzschild ISCO frequency $f_{\rm ISCO}$ \cite{Cutler_1994}, instead of the Kerr ISCO frequency, as the upper cut-off frequency of the Fisher analysis. We note that for (sub-)solar-mass superspinars with aligned spins, the corresponding ISCO frequencies are typically of order $\mathcal{O}(10)~\mathrm{kHz}$ \cite{Uchikata_2021}, well above the sensitive frequency range of the detector networks we are interested in.
For each network, we examine:
\begin{itemize}
    \item The distribution of SNR across the simulated population.
    \item The distribution of network SNR, the measurement uncertainties in the intrinsic binary parameters (chirp mass, symmetric mass ratio, and the aligned spin components $\chi_{1z}$ and $\chi_{2z}$), and the sky-localization performance.
\end{itemize}

In this work we include the effect of Earth's rotation in the detector response by enabling the corresponding option in GWBENCH \cite{Borhanian_2021}, as the signals considered are sufficiently long that this effect may impact parameter estimation. For a frequency-domain inspiral waveform, the stationary-phase approximation associates different frequencies with different times during the inspiral. Through the option given in GWBENCH, the detector antenna-pattern response accounts for the corresponding change in detector orientation as Earth rotates.

\subsection{\label{subsec:waveform_tf2} Gravitational Waveform Model}

We model GW signals from (sub-)solar-mass collapsed binaries using the TaylorF2 (TF2) frequency-domain inspiral approximant. This model describes the inspiral phase using the stationary phase approximation applied to the post-Newtonian expansion of the binary dynamics \cite{Sathyaprakash_1991,Damour_1998,Arun_2005}.

TF2 provides an accurate description of inspiral signals for collapsed binaries with aligned/anti-aligned spins in the frequency band of ground-based detectors. Since our analysis is restricted to the inspiral regime, TF2 is well suited for modeling the GW signals considered in this work. Furthermore, the post-Newtonian phase expansion remains mathematically well defined when the dimensionless spin parameter is extended beyond unity \cite{Uchikata_2021}.  We therefore employ TF2 as a phenomenological framework to investigate the sensitivity of inspiral-phase GW observations to apparent Kerr-bound violations. This should not be interpreted as a physically complete or self-consistent waveform model for superspinars, but rather as an exploratory tool for assessing the measurability of phenomenological departures from the Kerr bound within the inspiral regime.

Additionally, in parameter estimation studies we usually assume the object as a black hole, and limit the spin parameter as $|\chi| \leq1$ \cite{Abbott_2016,Abbott_2017,Abbott_2017_a,Abbott_2017_b,Abbott_2019,Abbott_2020_b,Abbott_2020_c}. Because of this, it is possible to misidentify a superspinar as a highly spinning, or extremal black hole. There are some detections having spin very close to Kerr bound \cite{Abbott_2020_c, Abbott_2020_d, Abbott_2021}. In addition to that, \cite{Biscoveanu_2021} shows that a different analysis method can lead to the possibility of existence of extreme black holes. Also, because of our pre-assumption on Kerr bound, there is no proper waveform model for binary merger of superspinars, especially in post-inspiral phase, we use TF2 waveform, which has no restriction on waveform spin value. Estimating the parameters of binaries with spins greater than 1 is discussed in \cite{Uchikata_2021, Broeck_2007, Wade_2013}.
Note that the absence of an event horizon (in case of superspinar) can, in principle, modify the inspiral waveform through the absence of horizon absorption effects. However, the standard TF2 waveform model employed in this work does not explicitly incorporate horizon absorption terms \cite{Favata}. Since these contributions enter at relatively high post-Newtonian order, our present analysis focuses on the dominant inspiral modifications arising from the superspinning nature of the collapsed objects. Our results should therefore be interpreted as sensitivity forecasts conditional on the phenomenological assumption that the leading post-Newtonian inspiral phasing can be analytically continued to \(\chi>1\), rather than as predictions for the waveform of a specific superspinar model.

\section{\label{sec:Results} Results}

In this section, we estimate errors on the set of $\lambda$ parameters considering the current and next-generation detector network, following the Fisher formalism described above. Our primary objective is to quantify the improvement in spin measurement precision and consequently the detectability of potential Kerr-bound violations when transitioning from current second-generation detector networks to third-generation configurations.

\subsection{\label{subsec:pararesults} Measurement uncertainties of various binary parameters}

\begin{figure*}[ht]
\centering
\begin{subfigure}[t]{0.495\textwidth}
    \centering
    \includegraphics[width=\textwidth]{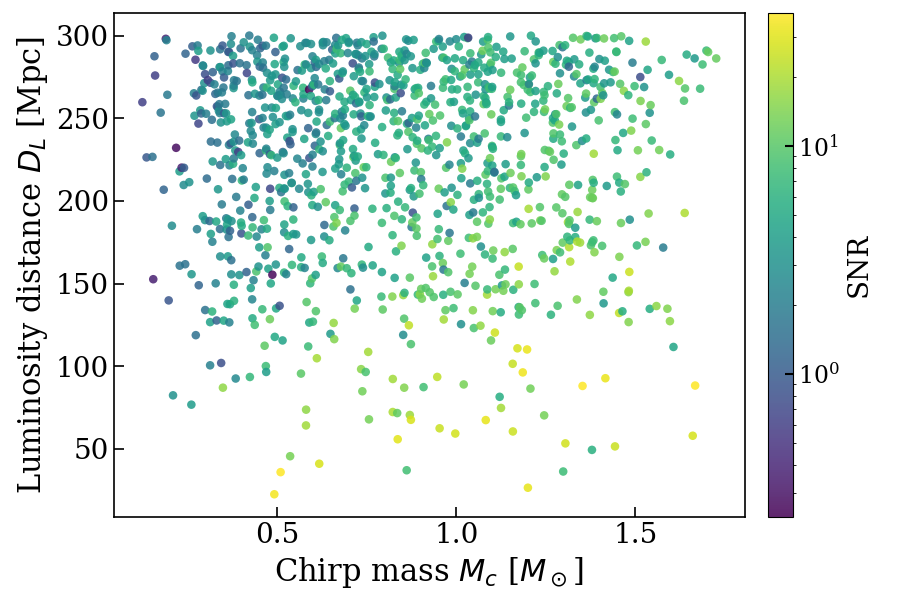}
    \caption{HL network}
    \label{fig:sub-a}
\end{subfigure}
\hfill
\begin{subfigure}[t]{0.495\textwidth}
    \centering
    \includegraphics[width=\textwidth]{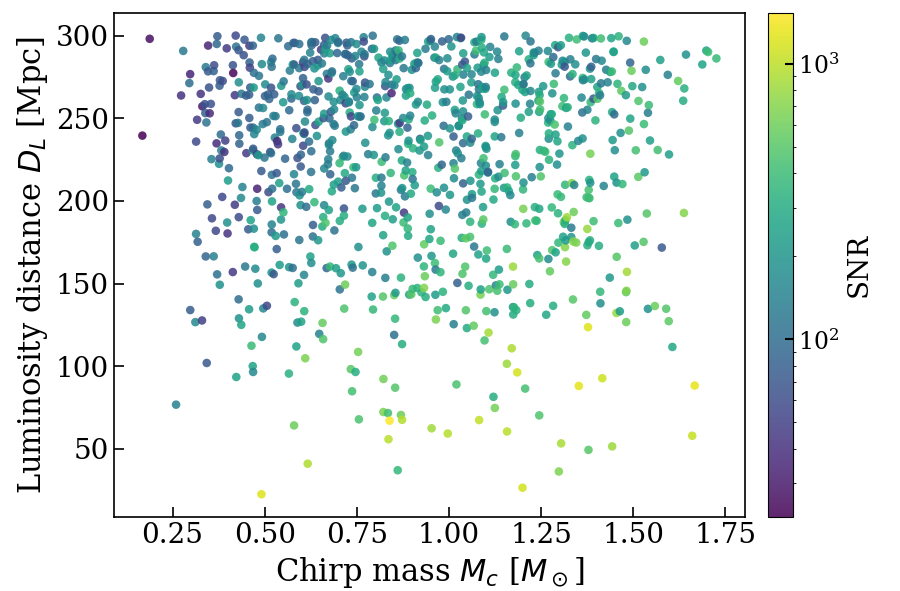}
    \caption{4040ET network}
    \label{fig:sub-b}
\end{subfigure}
\caption{Scatter plot of the network SNR as a function of chirp mass \(\mathcal{M}_c\) and luminosity distance \(D_L\) for the HL and 4040ET networks. The distribution reflects the combined effects of intrinsic binary parameters, source distance and orientation, detector response, and the frequency range contributing to the inspiral SNR. The SNR decreases with distance, while the apparent dependence on chirp mass should not be interpreted as an intrinsic fixed-parameter scaling. For more details, see section~\ref{subsec:pararesults}.}%
\label{fig:snrscatter}
\end{figure*}

Figure \ref{fig:snrscatter} shows the distribution of network SNR as a function of chirp mass \(\mathcal{M}_c\) and luminosity distance \(D_L\) for the simulated (sub -)solar-mass binary population observed by the HL (panel a) and 4040ET (panel b) networks. The figure represents the combined effect of the intrinsic binary parameters, luminosity distance, source orientation, sky location, detector response, and the frequency range contributing to the signal, and therefore should not be interpreted as an isolated SNR-chirp mass scaling at fixed source and detector parameters. The SNR decreases with increasing luminosity distance, as expected, while the apparent dependence on chirp mass reflects the combined dependence of the waveform amplitude and the frequency interval over which the inspiral contributes to the SNR. In particular, the longer duration of low-mass inspirals provides more accumulated phase information, but does not by itself imply a larger SNR.

The 4040ET population extends to substantially higher SNRs and chirp masses than the HL population. This reflects the higher sensitivity and broader frequency bandwidth of the third-generation network. In the present population study, the SNR distribution is used to characterize the detectable population and to select systems for subsequent parameter estimation analysis. A restricted analysis at fixed distance, orientation, and detector configuration is used separately to verify the expected increase of SNR on chirp mass. At fixed source distance and orientation, a separate consistency calculation confirms the expected increase of SNR with chirp mass using the TF2 waveform.

\begin{figure*}[ht]
\centering
\begin{subfigure}[b]{0.495\textwidth}
    \centering
    \includegraphics[width=\textwidth]{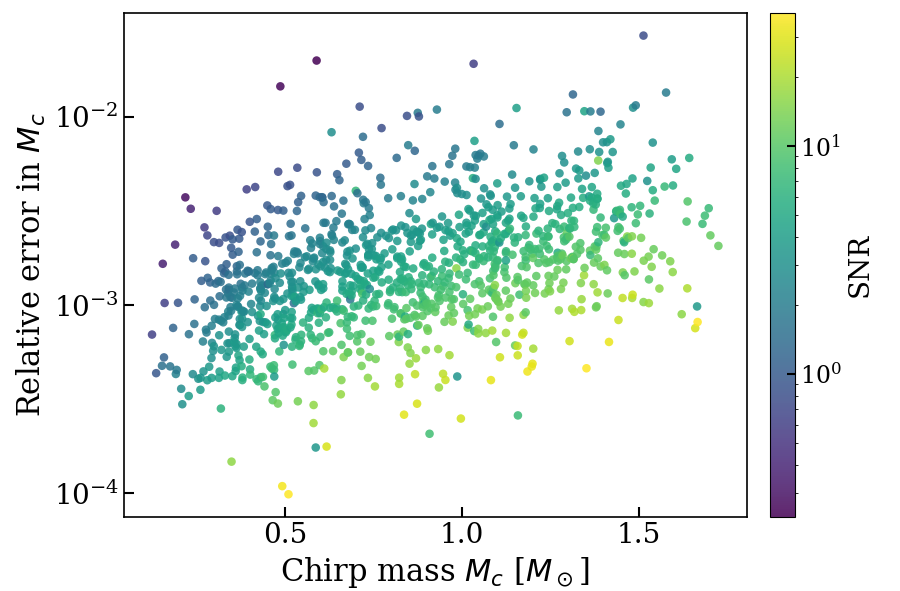}
    \caption{\centering HL network}
    \label{fig:sub-a}
\end{subfigure}
\begin{subfigure}[b]{0.495\textwidth}
    \centering
    \includegraphics[width=\textwidth]{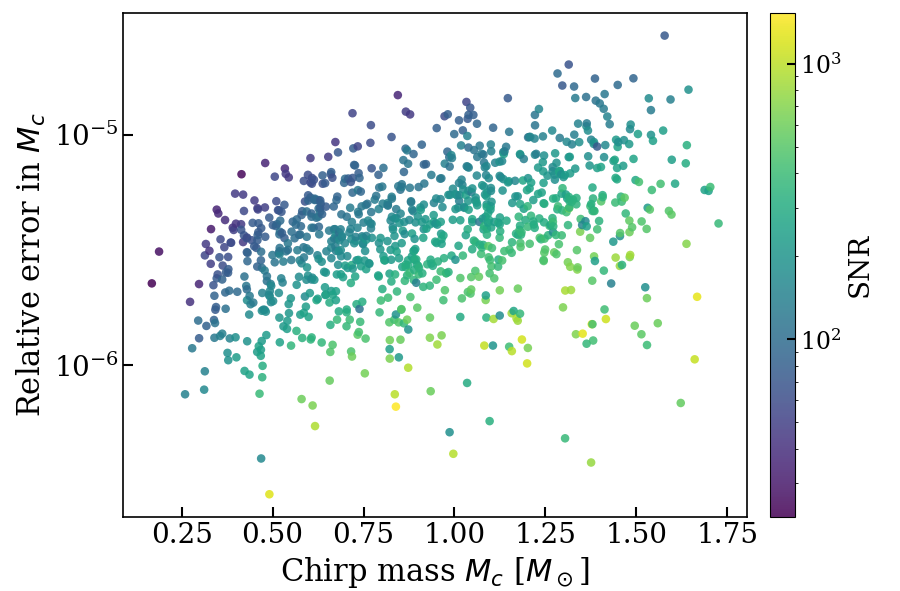}
    \caption{\centering 4040ET network}
    \label{fig:sub-b}
\end{subfigure}
\caption{Scatter plot of uncertainty in chirp mass with SNR for HL and 4040ET networks. For further details, see section \ref{subsec:pararesults}.}
\label{fig:errMc}
\end{figure*}

Figure~\ref{fig:errMc} shows the relative error in the chirp mass, $\Delta \mathcal{M}_c/\mathcal{M}_c$, for the HL network (panel a) and the 4040ET network (panel b). As expected, the third-generation detector network generally yields smaller relative errors, reflecting its greater sensitivity to GW signals. The distributions show a trend of error getting smaller as SNR increases, as expected.

\begin{figure*}[ht]
\centering
\begin{subfigure}[b]{0.495\textwidth}
    \centering
    \includegraphics[width=\textwidth]{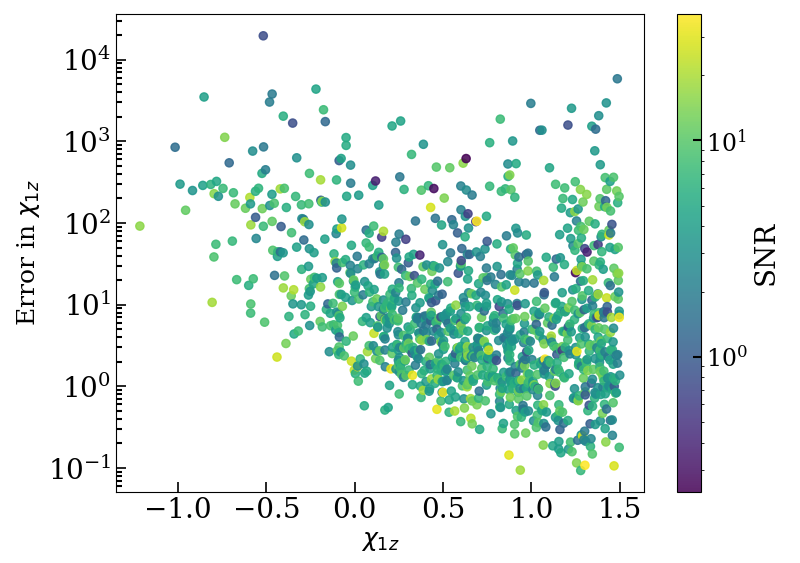}
    \caption{ \centering}
    \label{fig:sub-a}
\end{subfigure}
\begin{subfigure}[b]{0.495\textwidth}
    \centering
    \includegraphics[width=\textwidth]{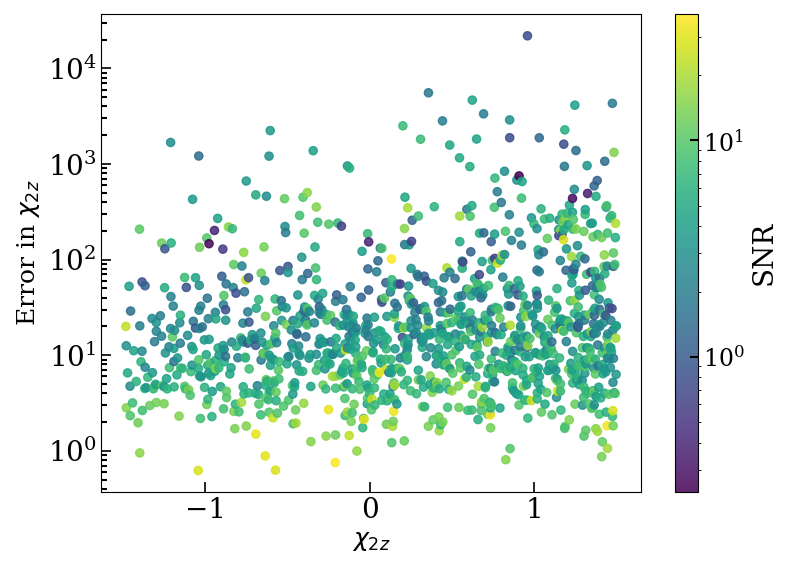}
    \caption{ \centering}
    \label{fig:sub-b}
\end{subfigure}
\caption{Uncertainties in the aligned component spins for the HL network. The individual-spin uncertainties exhibit a broad distribution, with some systems yielding very large uncertainties. This reflects the strong degeneracy between the two component spins in the inspiral waveform and the corresponding limitations of the Fisher approximation for poorly constrained systems. For more details see section \ref{subsec:pararesults}.}
\label{fig:errspin_1}
\end{figure*}

Figure~\ref{fig:errspin_1} shows the uncertainties in the measurements of the component spin parameters $\chi_{1z}$ and $\chi_{2z}$ for the HL network. While a fraction of the systems exhibit very small uncertainties, a substantial number of binaries show relatively large spin uncertainties. It is also evident that $\chi_{2z}$ is generally more poorly constrained than $\chi_{1z}$.

This behavior arises because the leading spin term in the inspiral phase is mass-weighted, so the spin of the lighter body ($m_2$) enters with a smaller weight and $\chi_{2z}$ is therefore the harder of the two to measure. The waveform is primarily sensitive to a single mass-weighted combination of the two spins, which is measured much better than either component, while the orthogonal combination is only weakly constrained. This near-degeneracy produces strong correlations between $\chi_{1z}$ and $\chi_{2z}$ and large uncertainties on the individual spins.

The spin uncertainties span several orders of magnitude, reaching $O(10^4)$ for a few systems. In the post-Newtonian expansion of the phase, the leading spin effects enter at 1.5PN order, well above the Newtonian chirp-mass term \cite{Cutler_1994, Poisson_1995}. As a result, the derivatives of the waveform with respect to the spin parameters are small, which leads to weak Fisher information and large uncertainties. In addition, the well-known degeneracy between the spin and mass parameters further degrades the measurement of the individual component spins. For many systems the Fisher matrix therefore predicts uncertainties far larger than the prior range ($|\chi| \le 1.5$). Such values should be read as showing that the spin is effectively unconstrained, not as a meaningful measurement.

\begin{figure*}[ht]
\centering
\begin{subfigure}[b]{0.495\textwidth}
    \centering
    \includegraphics[width=\textwidth]{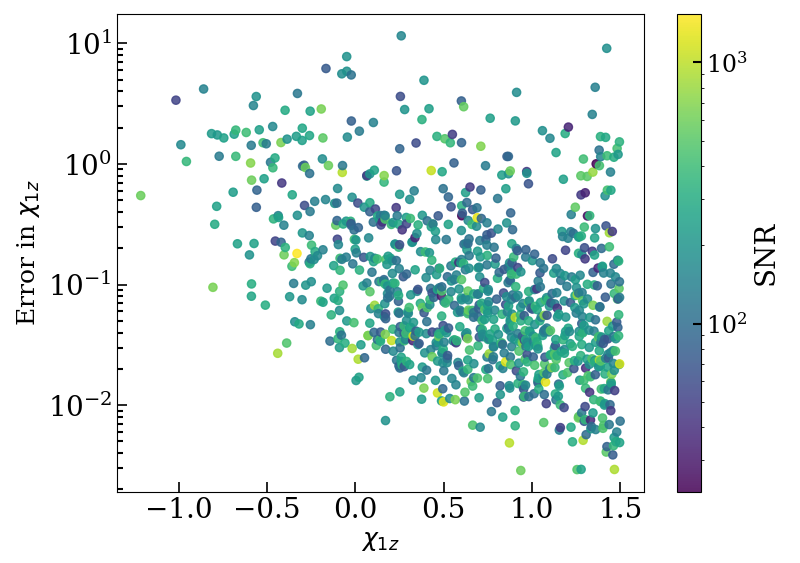}
    \caption{ \centering}
    \label{fig:sub-a}
\end{subfigure}
\begin{subfigure}[b]{0.495\textwidth}
    \centering
    \includegraphics[width=\textwidth]{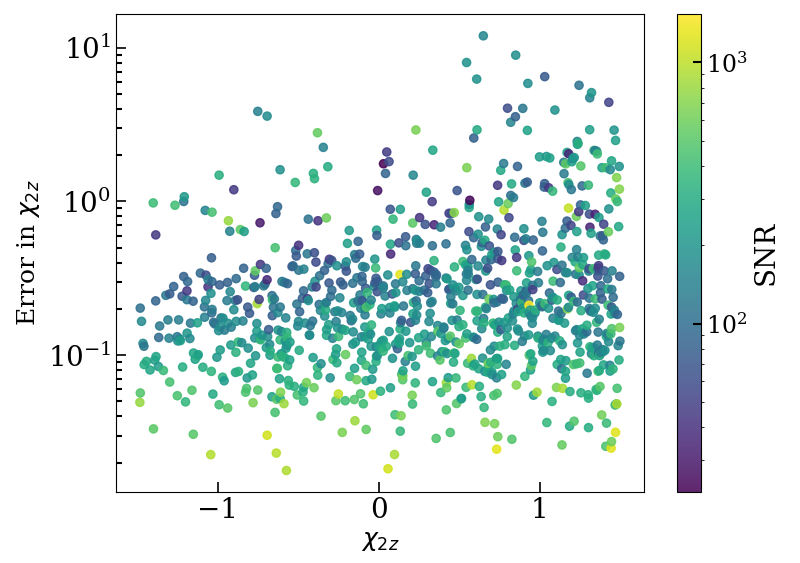}
    \caption{ \centering}
    \label{fig:sub-b}
\end{subfigure}
\caption{Same as figure \ref{fig:errspin_1} but for the 4040ET network. The substantially narrower distribution demonstrates the improved constraints on the individual spins afforded by the third-generation network, although a subset of systems remains poorly constrained. For more details see section \ref{subsec:pararesults}.}
\label{fig:errspin_2}
\end{figure*}

Figure~\ref{fig:errspin_2} shows the corresponding spin constraints for the 4040ET network. The uncertainties are significantly smaller than those obtained with the HL network, with most systems having uncertainties below $\sim 0.1$. Nevertheless, the relative behavior remains similar, with $\chi_{2z}$ still less well constrained than $\chi_{1z}$.

It is evident that third-generation detectors can measure the dimensionless spin parameter with comparatively high precision. For the systems considered here, with a masses range between $0.1~M_\odot$ and $2~M_\odot$ and the dimensionless spin projections along the orbital angular momentum in a range of $-1.5$ to $1.5$, the median uncertainties are $\Delta\chi_{1z}\simeq 0.07$ and $\Delta\chi_{2z}\simeq 0.2$, and the best-measured systems reach $\Delta\chi_{1z}\sim 3\times10^{-3}$, while about $9\%$ of systems remain essentially unconstrained ($\Delta\chi>1$). Table~\ref{table4} lists examples of the best-measured systems for the 4040ET network. These examples show the configurations for which the Fisher analysis gives the smallest marginalized spin uncertainties.

We note that the individual spin parameters exhibit a very strong anti-correlation, with median correlation coefficients of $-0.9987$ for the HL network and $-0.9996$ for the 4040ET network (computed from the marginalized covariance matrix for a random subsample of 150 converged systems per network).

Nevertheless, the marginalized uncertainties obtained with the third-generation network are smaller than those of the HL network by roughly two orders of magnitude in the median, which demonstrates the improved spin-measurement capability provided by the higher sensitivity and the extended low-frequency bandwidth. These values indicate the precision with which the component spins could be measured, under the adopted waveform and source assumptions, if such highly spinning objects exist. However, Fisher-matrix uncertainties should not by themselves be interpreted as evidence for a Kerr-bound violation, since for $|\chi|>1$ the TF2 phase is an extrapolation of the post-Newtonian expressions.

\begin{table}[h]\centering
\begin{tabular}{cccc}\hline
$\chi_{1z}$ & $\chi_{2z}$ & $\Delta\chi_{1z}$ & $\Delta\chi_{2z}$ \\ \hline
0.9367 & 1.4450 & $2.87\times10^{-3}$ & $2.74\times10^{-2}$ \\
1.2553 & 0.8264 & $2.92\times10^{-3}$ & $2.84\times10^{-2}$ \\
1.4669 & -0.5734 & $2.93\times10^{-3}$ & $1.78\times10^{-2}$ \\
1.2784 & 1.4750 & $2.93\times10^{-3}$ & $4.82\times10^{-2}$ \\
1.4576 & -0.6837 & $3.87\times10^{-3}$ & $1.47\times10^{-1}$ \\
1.4205 & -1.3987 & $4.08\times10^{-3}$ & $3.32\times10^{-2}$ \\
\hline\end{tabular}
\caption{Example binary configurations that yield particularly precise spin measurements with the 4040ET network. The resulting uncertainties illustrate the potential spin-measurement precision achievable with third-generation detectors. See section~\ref{subsec:pararesults} for further discussion.}
\label{table4}
\end{table}

\subsection{\label{subsec:skyloc}Sky-localization performance}

Determining the sky location of the source is an important aspect of GW observations. Accurate sky localization enables the identification of the host galaxy and facilitates potential multi-messenger follow-up observations. Associating a GW event with its astrophysical environment can therefore provide valuable clues about the origin of the source. It is thus useful to examine the sky-localization capability of current and future detector networks for such systems. The improved sky localization enables more targeted searches for electromagnetic or high-energy counterparts. If Kerr superspinars possess distinctive high-energy signatures, accurate GW localization could therefore facilitate complementary tests of the Kerr superspinar interpretation.

We restrict the sky-localization analysis to detectable systems with \(\rho\ge8\). We then calculate the marginalized 90$\%$ credible sky area. To illustrate the sky-localization performance, we list representative high-SNR systems in Tables \ref{table2} and \ref{table3}. The HL network (Table \ref{table2}), consisting of only two detectors, provides relatively poor sky localization even for high-SNR events. In contrast, the 4040ET network (Table \ref{table3}) benefits from both its three-detector geometry and the extended low-frequency sensitivity, yielding substantially smaller sky-localization areas for the systems considered here. The values should be regarded as representative Fisher-matrix estimates rather than definitive localization predictions, since the Fisher matrices can become highly ill-conditioned for individual systems.

Although a larger SNR generally improves localization, the sky-localization area also depends on how effectively the detector network constrains the source position. In particular, differences in signal arrival times between detectors provide information about the source direction, while the network geometry determines how these timing constraints intersect on the sky. Consequently, a high-SNR event does not necessarily have the smallest localization area, especially for a two-detector network such as HL.
Consequently, a relatively nearby source does not necessarily yield the smallest localization area, even when its SNR is high. This effect is particularly relevant for the HL network, whose two-detector geometry limits its ability to constrain the source position on the sky. 

\begin{table}[h]\centering
\begin{tabular}{ccccccc}\hline
$m_1$ & $m_2$ & $\chi_{1z}$ & $\chi_{2z}$ & $D_L$ & SNR & $\Delta\Omega_{90}$ \\
($M_\odot$) & ($M_\odot$) & & & (Mpc) & & (deg$^2$) \\ \hline
1.7587 & 1.3765 & 1.0767 & 0.7318 & 88.0879 & 38.4641 & 285.92 \\
1.0549 & 0.3438 & 1.3025 & -0.2102 & 35.9669 & 38.3832 & 692.19 \\
1.9555 & 1.8754 & 0.6904 & 0.1318 & 88.3317 & 38.1355 & 7469.7 \\
0.8204 & 0.3981 & 0.4997 & 1.4426 & 22.5862 & 35.1314 & 2163.5 \\
1.5292 & 1.2158 & 1.4965 & 0.0574 & 96.2827 & 34.1021 & 7384.4 \\
\hline\end{tabular}
\caption{Representative high-SNR detections of (sub-)solar-mass binary systems in the HL network. Even for relatively strong signals, the sky-localization uncertainty remains large, illustrating the limited localization capability of the two-detector network. For more discussions, see section \ref{subsec:skyloc}.}
\label{table2}
\end{table}

\begin{table}[h]\centering
\begin{tabular}{ccccccc}\hline
$m_1$ & $m_2$ & $\chi_{1z}$ & $\chi_{2z}$ & $D_L$ & SNR & $\Delta\Omega_{90}$ \\
($M_\odot$) & ($M_\odot$) & & & (Mpc) & & (deg$^2$) \\ \hline
1.0548 & 0.8820 & -0.3304 & 0.9326 & 67.0818 & 1531.4780 & 0.0047486 \\
1.9555 & 1.8754 & 0.6904 & 0.1318 & 88.3317 & 1319.8198 & 0.0027662 \\
1.7587 & 1.3765 & 1.0767 & 0.7318 & 88.0879 & 1279.0125 & 0.0060859 \\
1.8781 & 1.3407 & 0.8511 & 1.4697 & 123.6496 & 1247.6399 & 0.0082603 \\
0.8204 & 0.3981 & 0.4997 & 1.4426 & 22.5862 & 1243.3055 & 0.03886 \\
\hline\end{tabular}
\caption{Representative high-SNR detections of (sub-)solar-mass binary systems in the 4040ET network. The corresponding sky-localization areas are substantially smaller than those obtained with the HL network, illustrating the improved localization capability of the third-generation network. The 4040ET network yields substantially smaller sky-localization areas than the HL network for the representative systems shown. For further details, see section \ref{subsec:skyloc}.}
\label{table3}
\end{table}

\section{Astrophysical Binary Formation scenarios and Expected Merger Rates \label{sec:formation-scenario}}

While the parameter estimation results presented above quantify the observability of (sub-)solar-mass binaries, their astrophysical relevance depends on the existence of viable binary formation channels and sufficiently large merger rates. The formation of (sub-)solar-mass black holes and superspinars remains uncertain and may proceed through a variety of primordial and astrophysical mechanisms, as discussed in section \ref{sec:SSMBH_NS}. In this section, we consider several representative binary-formation scenarios and estimate the corresponding formation-rate densities.
We adopt optimistic benchmark populations because no theoretically well-established superspinar abundance is currently available. The resulting rates should therefore be interpreted as illustrative upper-limit estimates.

\subsection{Primordial Formation Channel\label{sec:5.1}}

Let us first focus on primordial collapsed binaries that form directly in the early Universe through the gravitational decoupling of nearby PBHs during the radiation-dominated epoch \cite{Sasaki_2016}. Their present-day merger rate is governed primarily by the distribution of merger delay times rather than by the efficiency of dynamical binary assembly. For an  equal-mass primordial binary population, the merger rate density ($\mathcal{R}$) can be approximated as \cite{raidal_2019,vaskonen_2020,stasenko_2023}

$$\mathcal{R} \sim 3.1 \times 10^6~S~f_{\rm PBH}^{53/37}~\left(\frac{m}{M_\odot}\right)^{-32/37}~\rm Gpc^{-3}~yr^{-1},$$
where, $S \sim 0.24 \left(1+2.3 \sigma_{\rm M}^2/f_{\rm PBH}^2\right)^{-21/74}$ is the suppression factor due to PBH interactions during binary formation, and $\sigma_{\rm M} \sim 0.005$ is the variance of matter fluctuations at the moment of PBH pair formation \cite{stasenko_2023}. $f_{\rm PBH}$ is the fraction of PBHs in dark matter \cite{stasenko_2023} and $m$ is the mass of each PBH in the binary system.

The merger rate depends sensitively on the values of $f_{\rm PBH}$. To obtain an illustrative estimate, we adopt $f_{\rm PBH}\sim 0.1$ and $0.01$ following \cite{stasenko_2023}, consistent with the upper limits derived from the LVK O3 search \cite{lvk_2023} for (sub-)solar-mass black hole binaries, which constrain $f_{\rm PBH}\lesssim 0.6$ around $m \sim 0.3 \rm M_\odot$ and  $f_{\rm PBH}\lesssim 0.09$ for $m \sim 1 \rm M_\odot$ for a monochromatic PBH mass distribution \cite{lvk_2023}. $f_{\rm PBH} \sim 0.1$ yields a merger rate of $4.8 \times 10^3-6.4 \times 10^4~\rm Gpc^{-3} yr^{-1}$ for PBH mass range of $0.1-2~\rm M_\odot$, while $f_{\rm PBH} \sim 0.01$ gives a merger rate of $1.8 \times 10^2-2.4 \times 10^3~\rm Gpc^{-3} yr^{-1}$. These estimates are broadly consistent with the rates inferred from recent candidate-event studies \cite{Prunier_2024, lvk_2023}.

\subsection{Dynamical Formation in Dark Matter Halos}

Among the possible formation channels, GW capture during close encounters provides a simple and generic mechanism for forming binaries from initially unbound collapsed objects. This process has been studied extensively in the context of compact objects in dense stellar systems and dark-matter halos \cite{QuinlanShapiro1989,OLeary2009,Bird2016}. In such encounters, two compact objects lose orbital energy through gravitational radiation during a close hyperbolic passage, potentially forming a bound binary that subsequently merges.

When two initially unbound compact objects undergo a close hyperbolic encounter, sufficient orbital energy can be radiated in GWs to form a bound binary. In such a case, the capture cross section can be written as \cite{QuinlanShapiro1989},

\begin{equation}
    \sigma_{\rm cap} = 2\pi \left(\frac{85\pi}{6\sqrt{2}}\right)^{\frac{2}{7}} \frac{G^2 \left(m_1+m_2\right)^{\frac{10}{7}} m_1^{\frac{2}{7}} m_2^{\frac{2}{7}}}{c^{\frac{10}{7}} v^{\frac{18}{7}}},
    \label{Eq:capture_cross_section}
\end{equation}
where $m_1$ and $m_2$ are component masses and $v$ is the relative velocity at infinity. For the purpose of calculation, we consider equal-mass binaries with  $m_1=m_2=0.5 M_\odot$, and adopt a characteristic halo velocity dispersion $v \sim 100~\rm km~s^{-1}$, consistent with the circular velocity of a Milky Way–mass dark matter halo \cite{Klypin_2002, Xue_2008}. Substituting these values, one obtains
$\sigma_{\rm cap} \sim \rm~ 10^{-35}~Gpc^2$.
We model a Milky Way–like dark matter halo with virial radius $R \sim 10^{-4} \rm~ Gpc$, consistent with observational and dynamical estimates for the Milky Way halo \cite{Klypin_2002}, and, hence the halo volume ($V$) is $V \sim \rm~ 10^{-12}~Gpc^3$.

Milky Way is expected to contain $N \sim 10^8-10^9$ neutron stars \cite{Sartore_2010, Reed_2021} and $\sim 10^9 - 10^{10}$ white dwarfs (scaling from the white dwarf number density of $5 \times 10^{-3}~\rm pc^{-3}$ \cite{Fontaine2001}), so even a small fraction (say, $f_{\rm tr}$) undergoing transmutation \cite{mcd2012, Chakraborty_2024, acb1} could populate the halo with a large number of low-mass collapsed objects.  This value should be interpreted as an illustrative benchmark rather than a constraint-driven estimate, intended to explore the maximum merger rate that could arise from such a population.
Thus, the corresponding number density is,
   $ n = f_{\rm tr}~N/V \sim \rm~ 10^{20}~Gpc^{-3}$
for $f_{\rm tr} \sim 0.01$ and $N \sim 10^{10}$.
Thus, the binary formation rate ($\mathcal{R}_b$) density for a halo is,
\begin{equation}
    \mathcal{R}_b = \frac{1}{2} n^2 \sigma_{\rm cap} v\approx\rm~ 10^{-7}~Gpc^{-3}~yr^{-1},
\end{equation}
where the factor $1/2$ avoids double counting of pairs \cite{Binney2008}. 
For the representative \(0.5+0.5\,M_\odot\) system with
\(v \sim 100\,{\rm km\,s^{-1}}\), the captured binaries are born with very high
eccentricities, \(e_0\gtrsim0.99\). Using the equations for
GW-driven orbital evolution, we find that systems captured with
\(r_p\lesssim0.9\,r_{p,\max}\) merge within approximately a few weeks,
with substantially shorter merger times for deeper encounters. Here,
\(r_p\) is the pericenter distance of the initial encounter and
\(r_{p,\max}\) is the maximum pericenter distance for which the energy
radiated in GWs exceeds the initial orbital energy at infinity.
Consequently, except for marginal captures with
\(r_p\rightarrow r_{p,\max}\), the delay between capture and merger is
negligible on astrophysical timescales. We therefore approximate the
GW-capture rate as the merger rate in this order-of-magnitude estimate.

\subsection{Formation Scenarios at the Galactic Center\label{sec:5.3}}

We now examine the possibility that (sub-)solar-mass  binaries form and merge in the Galactic Center. If such mergers occurred within the central few kiloparsecs of the Galaxy, their detectability by current GW detectors would be essentially guaranteed. The distance to the Galactic Center is $D_{\rm L} \sim 8 \rm~ kpc$. Since GW SNR scales $\rho \propto \mathcal{M}_c^{5/6}/D_L$ \cite{Cutler_1994, Finn_1992}, a (sub-)solar-mass binary at $8 \rm~ kpc$ would produce an enormous SNR in the aLIGO network. Scaling from typical binary neutron star detections at $\sim 100 \rm~ Mpc$ with $\rho \sim 10$, one obtains, $\rho_{\rm GC} \approx 10^4 - 10^5$ where $\rho_{\rm GC}$ denotes the network SNR for a source located at the Galactic Center, even for current detector sensitivities. Such a signal would be unmistakable. The absence of such detections is consistent with a strongly suppressed formation rate of (sub-)solar-mass binaries in this environment.

Observational and theoretical studies suggest that a substantial
population of compact remnants may reside in the central region of the
Galaxy. Mass-segregation models predict that compact objects can
accumulate within the central \(0.1\)--\(1~{\rm pc}\) through dynamical
friction and stellar evolution \cite{Hopman_2006,Alexander_2009}.
Representative models yield populations of approximately
\(10^3\)--\(10^4\) compact remnants in this region
\cite{Hopman_2006,Chakraborty_2024,Calamida_2015,NASA2015,Kim_2018}.
As an optimistic benchmark, we take \(N=10^4\) potential progenitor
objects within an effective volume \(V=1~{\rm pc}^3
=10^{-27}~{\rm Gpc}^3\) and assume that a fraction
\(f_{\rm tr}=0.1\) undergoes transmutation into low-mass collapsed
objects. The resulting number density is
$ n=f_{\rm tr} N/V
   \sim 10^{30}~{\rm Gpc}^{-3}$.
The assumed transmutation fraction is not an observationally determined
value; it is adopted as an optimistic benchmark motivated by the
potentially high dark-matter density in the Galactic Center
\cite{Chakraborty_2024,Chakraborty_2024b,acb1,acb2}. We then examine
three possible dynamical formation channels: three-body encounters, exchange interactions with pre-existing binaries, and
two-body GW capture.

\subsubsection{Three-body Encounters}
In dense stellar systems, binaries can form dynamically through three-body encounters, in which two initially unbound objects become gravitationally bound by transferring excess kinetic energy to a third passing object. The third body carries away the excess energy, allowing the remaining pair to form a bound binary. This process is strongly enhanced by high number densities and low velocity dispersions. For an equal-mass population, the binary formation rate per unit volume ($\mathcal{R}_b$) can be approximated as \cite{GoodmanHut1993,Dawoo2017},
$$
\mathcal{R}_b
\simeq
0.75\,\frac{G^5m^5n^3}{v^9} \sim 1~{\rm Gpc^{-3}\,yr^{-1}}
$$
for \(m\sim1M_\odot\).
This represents the three-body binary formation rate density; the corresponding merger rate requires an additional factor accounting for the fraction of dynamically formed binaries that subsequently merge within the relevant timescale. The merger rate ($\mathcal{R}$) is therefore \(\mathcal{R}=f_{\rm m}\mathcal{R}_b\), where \(f_{\rm m}\) is the fraction of dynamically formed binaries that merge within the relevant timescale. If the calculated post-formation merger times are much shorter than the characteristic timescale over which the binary population is replenished, then \(f_{\rm m}\sim1\), and the merger rate is of the same order as the formation rate.

\subsubsection{Binary Exchange Interactions}

Exchange interactions require the presence of a pre-existing binary population. However, binaries in the central parsec of the Galaxy are expected to be efficiently disrupted by dynamical encounters and tidal perturbations from the central supermassive black hole, Sgr A* \cite{Hills1988,Heggie1975}. The dynamical evolution of binaries in such environments is governed by Heggie’s law \cite{Heggie1975}, which states that hard binaries tend to harden while soft binaries are disrupted, together with strong tidal perturbations induced by the central potential \cite{Hills1988}. Direct $N$-body simulations of the Galactic Center find that the binary fraction in the inner parsec can remain at the level of a few percent after several Gyr, with soft binaries preferentially destroyed \cite{Panamarev2019}. Thus, although a population of hard binaries can survive in the central parsec, the low binary fraction limits the number of potential targets for exchange interactions.

For a hard binary of semi-major axis $a \leq 0.1$ AU, the gravitationally focused exchange cross section is approximately,
$$\sigma_{\rm ex}=2\pi G \left(M_{\rm bin}+m\right)\frac{a}{v^2}\sim \rm~ 10^{-30}~Gpc^{2},$$
for representative stellar masses $m \sim 1~ M_{\odot}$. Here $\sigma_{\rm ex}$ is the binary exchange cross-section and $M_{\rm bin}$ is the total mass of binary system \cite{Heggie1975}. Since the binary fraction ($N_{\rm bin}$) of the compact-object population in the Galactic Center is poorly constrained, we adopt the binary fraction \(f_{\rm bin}=1\%\) as an illustrative benchmark, that gives
$N_{\rm bin} = \frac{1}{2}~f_{\rm bin}\times N\approx 50.$
For a pre-existing binary, the exchange interaction rate is given by $\Gamma_{\rm ex}=n\sigma_{\rm ex}v$. Thus, the exchange interaction rate per unit volume is,
$$\mathcal{R}_b\approx~ \Gamma_{\rm ex} \frac{N_{\rm bin}}{V} \sim ~10^{7}~\rm Gpc^{-3}~yr^{-1}.$$
The subsequent merger time of the binaries depends on their orbital parameters. For representative \(1\,M_\odot+1\,M_\odot\) binaries, the merger time can range from much shorter than the Hubble time for highly eccentric and/or compact binaries to longer than the Hubble time for wider, less eccentric systems. Since the present calculation does not specify the post-formation distributions of semi-major axis and eccentricity, we do not assign a unique merger delay time. Nevertheless, the merger rate cannot exceed the exchange interaction rate.

\subsubsection{Two-Body Gravitational wave Capture}

Two initially unbound compact objects may form a binary if sufficient orbital energy is radiated during a close encounter. The capture cross section was derived in \cite{QuinlanShapiro1989}, which is given by eq. (\ref{Eq:capture_cross_section}). Thus the binary formation rate per unit volume is obtained as,
$$\mathcal{R}_b\approx n^2\sigma_{\rm cap}v~\approx~10^{10}~\rm Gpc^{-3}~yr^{-1}.$$
For highly eccentric binaries, the calculated merger time is usually very less than the Hubble time. Hence, to order of magnitude, the capture rate can be identified with the merger rate.

\subsection{Summary of Formation Scenarios}
The results presented in sections \ref{sec:5.1}-\ref{sec:5.3} demonstrate that the merger rate of the transmuted collapsed objects depends on both the formation channel and the local astrophysical environment. In particular, the Galactic Centre exhibits a substantially enhanced merger rate owing to its higher dark-matter and stellar densities. Although the effective volume of the Galactic Centre is relatively small compared with that of the Galactic halo, the enhanced local densities can compensate for the smaller volume, making the Galactic Centre a potentially favourable environment for the formation and subsequent mergers of (sub-)solar-mass collapsed objects. A similar enhancement is obtained for the PBH channel, although the relative contributions of the different channels depend on their respective formation rates and parameter ranges.

These results indicate that dense central galactic environments, particularly the Galactic Centre, may provide more favourable conditions than the Galactic field for producing merger events involving (sub-)solar-mass collapsed objects. This motivates future GW searches for such systems, especially from dense galactic environments. The predicted merger rates should, however, be interpreted within the assumptions adopted for the dark-matter distribution, compact-object population, effective volume, and formation and merger efficiencies considered in sections \ref{sec:5.1}-\ref{sec:5.3}.

\section{\label{sec:Summary}Conclusions and Discussion}

In this work, we have investigated the detectability and parameter estimation prospects of (sub-)solar-mass collapsed binaries using current and next-
generation GW detector networks. Employing the
inspiral-only TF2 waveform model within a
Fisher-matrix framework, we have quantified the expected measurement precision for key binary parameters, including chirp mass, component spins, and sky
localization, while allowing for phenomenological extensions of the spin parameter beyond the Kerr bound.

Our results indicate that second-generation detectors such as aLIGO have a substantially smaller detection volume and parameter estimation reach for (sub-)solar-mass binaries than the third generation networks. While some nearby systems may yield detectable SNRs, the parameter estimation accuracy, particularly for spin and sky localization, remains relatively poor. Notably, sky-localization uncertainties for high-SNR systems typically span large areas, making host environment identification challenging.
In contrast, third-generation detector networks such as the ET and CE substantially improve the detectability and parameter estimation precision of these systems. Their improved strain sensitivity and extended low-frequency coverage provide access to a larger portion of the inspiral, resulting in higher SNRs and substantially tighter constraints on the binary parameters. In particular, the Fisher analysis indicates significantly improved sensitivity to the spin sector and much better sky localization compared with the current-generation network. However, the component spins exhibit strong anti-correlations, and the full Fisher matrices can become highly ill-conditioned for individual systems. Consequently, the small marginalized spin uncertainties obtained for favorable configurations should be interpreted as measures of potential parameter estimation precision rather than as standalone evidence for a violation of the Kerr bound.

While the detectability of such systems improves markedly with future detectors, their astrophysical formation must also be considered. We have specifically examined the possibility of (sub-)solar-mass binaries in the Galactic Centre, where the high dark-matter and stellar densities could enhance their formation rate. Such nearby systems would produce very large SNRs and should therefore be detectable even with current detectors if they occur in appreciable numbers. The absence of such detections may therefore constrain the abundance and formation efficiency of these systems. Our estimates, however, indicate that the binary formation rate of (sub-)solar-mass collapsed objects can be comparatively high in the Galactic Centre if the dark-matter-induced transmutation scenario is realized. This highlights the Galactic Centre as a potentially important environment for producing such binaries and motivates further investigation of their formation channels and observational signatures.

In conclusion, this study has investigated the potential of GW observations of (sub-)solar-mass binaries to probe spin parameters beyond the Kerr bound. By extending the TF2 inspiral waveform phenomenologically to allow \(|\chi_{1z},\chi_{2z}|>1\), we find that current and next-generation detector networks can provide meaningful constraints on the intrinsic parameters of such systems, with third-generation detectors offering substantially improved sensitivity and parameter estimation precision. Additionally, we emphasize that the \(\chi>1\) extension adopted in this work is phenomenological. Consequently, the reported parameter estimation accuracies should be interpreted as sensitivity forecasts conditional on the analytical continuation of the leading post-Newtonian inspiral phasing beyond the Kerr bound, rather than as predictions for a specific physical superspinar waveform. A definitive assessment of superspinar signatures will require physically consistent waveform models and Bayesian model comparisons.

\section{Acknowledgments}

Sruthy acknowledges the support of the DST-INSPIRE Fellowship, Govt. of India (Reference No: DST/INSPIRE/03/2023/003051, INSPIRE code: IF230417). Krishnendu is supported by STFC grant ST/Y00423X/1. Chandrachur acknowledges the support of Manipal Academy of Higher Education. N.~Uchikata acknowledges the support from JSPS KAKENHI Grant Number JP23K03381. Authors thank Poulami Dutta Roy for carefully reading the manuscript and providing useful comments. This document has LIGO preprint number {\texttt{P2600222}}.

\bibliographystyle{unsrt}
\bibliography{references}

\end{document}